\documentclass[preprint2,twoside]{hwo}
\graphicspath{{./}{figures/}}
\usepackage[printonlyused,nohyperlinks]{acronym}
\usepackage{lipsum}

\newcommand{\citeac}[2]{\aclu{#1} \citep[\ac{#1};][]{#2}}
\newcommand{\focusgroupleads}[1]{\noindent\textcolor{violet}{Focus Group Leads: {#1}}}
\newcommand{\focusgrouplead}[1]{\noindent\textcolor{violet}{Focus Group Lead: {#1}}}
\input{hwo.h}

\begin{document}

\title{\textbf{\LARGE Habitable World Discovery and Characterization: Coronagraph Concept of Operations and Data Post-Processing}}

\author{\textbf{\large Michael W. McElwain,$^{1}$ Dimitri Mawet,$^{2,3}$ Roser Juanola Parramon,$^{4,1,5}$ Kellen Lawson,$^{4,1,5}$ Hervé Le Coroller,$^{6}$ Christian Marois,$^{7}$ Max Millar-Blanchaer,$^{8}$ Bijan Nemati,$^{9}$ Susan F. Redmond,$^{2,3}$ Bin Ren,$^{10,11}$ Jean-Baptiste Ruffio,$^{12}$ Laurent Pueyo,$^{13}$ Christopher Stark,$^{1}$ \& Scott Will$^{1}$}}

\affil{$^{1}$\small\it NASA Goddard Space Flight Center, Greenbelt, Maryland, USA}
\affil{$^{2}$\small\it Department of Astronomy, California Institute of Technology, Pasadena, California, USA}
\affil{$^{3}$\small\it Jet Propulsion Laboratory, California Institute of Technology, Pasadena, California, USA}
\affil{$^{4}$\small\it Center for Space Sciences and Technology, University of Maryland, Baltimore County, Baltimore, Maryland, USA}
\affil{$^{5}$\small\it Center for Research and Exploration in Space Science and Technology, NASA-GSFC, Greenbelt, Maryland, USA}
\affil{$^{6}$\small\it Aix Marseille Université, CNRS, CNES, LAM, Marseille, France}
\affil{$^{7}$\small\it NRC Herzberg Astronomy and Astrophysics Research Centre, Victoria, British Columbia, Canada}
\affil{$^{8}$\small\it Department of Physics, University of California, Santa Barbara, Santa Barbara, California, USA}
\affil{$^{9}$\small\it Tellus1 Scientific, LLC, Huntsville, Alabama, USA}
\affil{$^{10}$\small\it Université Côte d’Azur, Observatoire de la Côte d’Azur, CNRS, Laboratoire Lagrange, Nice, France}
\affil{$^{11}$\small\it Max-Planck-Institut für Astronomie, Heidelberg, Germany}
\affil{$^{12}$\small\it Department of Astronomy \& Astrophysics, University of California, San Diego, La Jolla, California, USA}
\affil{$^{13}$\small\it Space Telescope Science Institute,  Baltimore, Maryland, USA}


\begin{abstract}
 The discovery and characterization of habitable worlds was the top scientific recommendation of the Astro2020 decadal survey and is a key objective of the \textit{Habitable Worlds Observatory}. Biosignature identification drives exceedingly challenging observations, which require raw contrasts of roughly 10$^{-10}$ contrast and ultimately, 1$\sigma$ photometric precision of roughly 3$\times~10^{-12}$ contrast. Despite significant advances for the \textit{Nancy Grace Roman Space Telescope}’s Coronagraph Instrument, technological gaps still exist in a wide range of technologies such as starlight suppression, deformable mirrors, wavefront control, low noise detectors, and high-contrast spectroscopy. Even with these new technologies matured, the \textit{Habitable Worlds Observatory} must carefully obtain the observations and rely on post-processing of the data to achieve its science objectives.

During the START and TAG efforts, a working group was convened to explore the Coronagraph Concept of Operations and Post Processing (COPP) in the context of the \textit{Habitable Worlds Observatory}. This COPP working group evaluated coronagraphic concept of operations to enable different post processing approaches, such as reference differential imaging and angular differential imaging, polarization differential imaging, orbital differential imaging, coherent differential imaging, spectral processing, and point-spread function subtraction algorithms that incorporate ancillary telemetry and data. Future integrated modeling simulations and testbed demonstrations are needed to determine the achievable post processing gains for each approach. We report a summary of this working group’s activities and findings, as well as an outlook for maturation of these techniques and infusion into the \textit{Habitable Worlds Observatory} technology portfolio.
  \\
  \\

\end{abstract}

\acresetall 

\section{Introduction}

A primary science goal of the \ac{hwo} will be to discover habitable worlds and characterize their spectral reflectance properties to search for signs of life, answering the question, ``Are we alone?''. While this scientific question was first asked by Democritus in the 5$^{th}$ century B.C., humanity now has the scientific knowledge and technical capability to build an observatory that answers this question. 

Observing habitable worlds requires a large, ultrastable telescope equipped with starlight suppression capable of measuring planet to star flux ratios of 10$^{-11}$. The residual optical misalignments and wavefront aberrations within the optical train will be sensed by metrology on the telescope and wavefront sensors within the coronagraph instrument. Control will be used to maintain alignment of the telescope segments using mechanical actuators and correct wavefront aberrations using tip/tilt and deformable mirror(s) within the coronagraph instrument. Nevertheless, the raw high-contrast scene will have residual uncorrected starlight, exozodical light, astrophysical backgrounds (e.g., background stars or galaxies), and detector noise that need to be differentiated from the planet signal.

\begin{figure*}[ht]
\begin{center}
\includegraphics[width=0.67\textwidth]{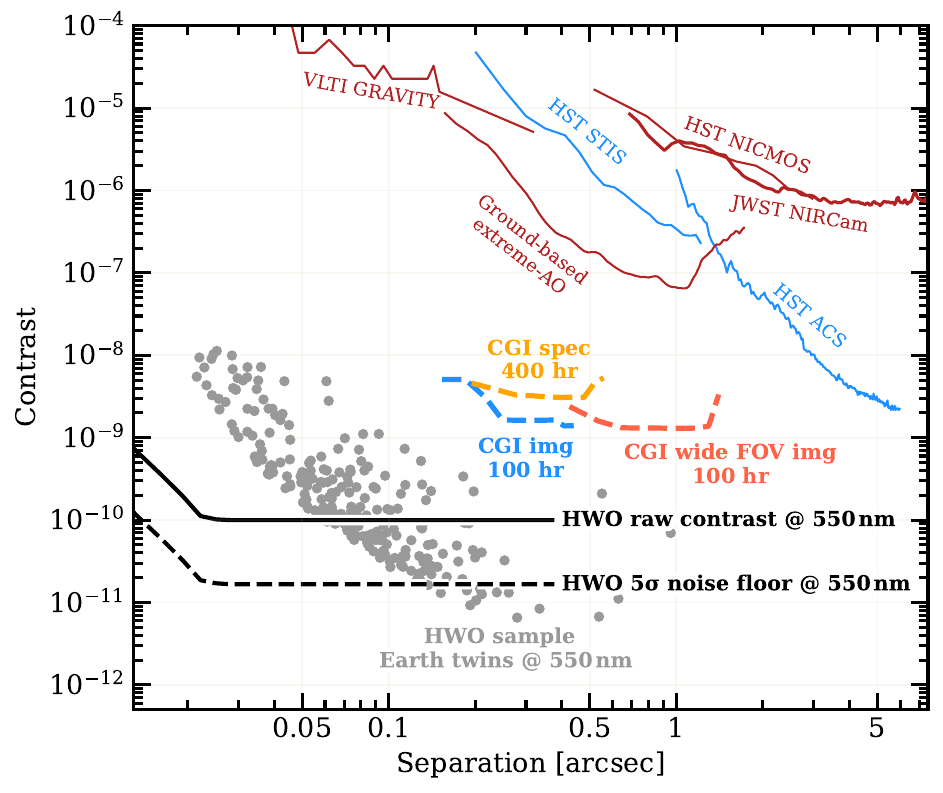}
\caption{\small While the \ac{ngrst} \ac{cgi} will work at far higher contrasts than existing facilities, habitable world discovery and characterization require operating more than an order of magnitude deeper. \Ac{hwo} sample Earth twins contrast at quadrature at 550~nm are plotted for reference. This plot made use of Vanessa Bailey's GitHub repository (\url{https://github.com/nasavbailey/DI-flux-ratio-plot}).
\label{fig:hwo_contrast_vs_others}
}
\end{center}
\end{figure*}

Post processing is the method by which the high-contrast data is analyzed to extract and analyze the planet and background signals within the scene. Post processing improves the discovery space and will be necessary to achieve the desired signal to noise (SNR) on the habitable worlds (Figure \ref{fig:hwo_contrast_vs_others}). There are numerous post processing methods that are promising for use on \ac{hwo} (see $\S$~\ref{sec:PostProcessingApproaches}). 

The observations needed to characterize habitable worlds are formidable but within reach. The \ac{hwo} design builds upon heritage from the large segmented telescope architecture from the \citeac{jwst}{McElwain+Feinberg+Perrin+etal_2023}, the ultrastable telescope architecture of the \citeac{ngrst}{Bolcar+Abel+Bartusek+etal_2023}, and the \ac{ngrst} \citeac{cgi}{Bailey+Bendek+Monacelli+etal_2023}. 


In 2023, the \ac{hwo} \ac{gomap} created a \ac{start} and NASA \ac{tag} with the charge to form and coordinate a series of groups whose collective activities would perform analyses and advance \ac{hwo}’s concept maturity. The \ac{hwo} Concept of Operations and Post Processing Working Group was established within the \ac{tag} structure and chaired by Dimitri Mawet and Michael McElwain, which reported through the \ac{hwo} \textit{Systems} Working Group. 

The Coronagraph Concept of Operations and Post Processing (COPP) Working Group formed focus groups for the coronagraph concept of operations and each of the post processing approaches briefly highlighted in \S~\ref{sec:PostProcessingApproaches}. Each of the focus groups were independently chaired and charged to develop white papers that included a description of the post processing approach with references to state of the art, the capabilities needed within the coronagraph instrument, the baseline needs for a concept of operations, the description of the simulated data needed with key parameters to vary, the analysis approach for simulated data, and an estimate level of effort to complete the analysis. The \ac{copp} white papers are publicly available on a \ac{hwo} repository (\url{https://bit.ly/HWO-COPP}) and several groups have papers that will appear in upcoming journals \citep{Ruffio+Steiger+Spohn+etal_2025,Redmond+Will+Sanchez+etal_2025,LeCoroller+Nowak+Hara+etal_2025}.

\section{Post Processing Approaches}\label{sec:PostProcessingApproaches}

There are numerous post processing options that can be pursued to aid in the starlight suppression and spectral characterization. Each approach was largely considered independently, but combinations of post processing approaches could yield better results. The HWO should evaluate each approach to determine which can be incorporated into the flight system to produce the best results. Some reduction methods must be considered from the outset of the project, during the instrument design phase. For instance,  Polarization Differential Imaging (PDI) algorithms require specialized hardware architecture, including optical and mechanical modifications as described in \S~2.2. Algorithms that use Kepler’s laws (see \S~2.3) to combine multiple observations taken at different epochs will gain efficiency if the wavefront can be stablized over long periods and if overhead times are minimized. Finally, it will be necessary to record telemetry (e.g., wavefront sensor) for certain \ac{psf} reconstruction methods (see \S~2.6).

\subsection{Reference Differential Imaging / Angular Differential Imaging}
\focusgroupleads{Roser Juanola Parramon, Christian Marois}

\Ac{rdi} is a high-contrast imaging post-processing technique designed to reduce stellar speckle noise by utilizing images acquired from other targets for PSF subtraction. Modern RDI algorithms employ singular value decomposition and \ac{pca} methods, such as \citeac{klip}{Soummer+Pueyo+Larkin_2012}, to constrain the reference image space and limit noise propagation. However, \ac{rdi} faces several fundamental limitations that impact its performance. Wavefront evolution between science and reference targets can reduce the post processed contrast gain, particularly as the temporal separation increases \citep{JuanolaParramon+Zimmerman+Pueyo+etal_2022}. Spectral differences between targets lead to residual speckle noise that scales chromatically and with bandwidth size, while stellar size discrepancies can result in differential speckles leaking through the coronagraph and degrade performance near the coronagraph inner working angle. Additional complications arise from star disk intensity asymmetries, coronagraph acquisition misalignments, and flux-dependent detector effects that compromise subtraction when images are normalized.

\Ac{adi} is a high-contrast imaging technique originally inspired by the \textit{Hubble Space Telescope}'s ``roll subtraction'' method and subsequently developed for ground-based telescopes in the mid-2000s at the Gemini North Telescope and the W.~M. Keck Observatory \citep{Marois+Lafreniere+Doyon+etal_2006}. The technique exploits field-of-view rotation to differentiate between quasi-static speckle patterns and astrophysical point sources, with images typically analyzed per $\lambda$/D annulus to optimize contrast at each angular separation. ADI can employ similar post-processing approaches to modern RDI referenced above.  For space observatories, operational constraints typically limit observations to two roll angles due to star reacquisition requirements and coronagraphic mask alignment challenges. ADI performance is fundamentally limited by several factors: wavefront stability variations during field rotation, deterministic speckle noise that may not achieve whitening through limited roll sequences, and partial self-subtraction of point sources at small angular separations. Additional complications include exozodiacal light structures that introduce extra noise in reconstructed reference PSF images, star reacquisition offsets that generate speckle noise, and stellar disk asymmetries or spot patterns that create uncorrelated intensity variations between roll angles, impacting both detectability thresholds and characterization precision for exoplanetary systems.

The evaluation of \ac{rdi} and \ac{adi} post-processing techniques requires comprehensive simulated observations similar to those developed for the \ac{ngrst} Observing Scenario 11 (NGRST OS11, \citealt{Krist+Steeves+Dube+etal_2023}), with enhanced modeling to accommodate the extreme contrast requirements anticipated for the \ac{hwo}. The simulated data framework should incorporate effects through the entire optical train, from the telescope to the detector, and astrophysical scene complexity to establish upper and lower limits for post-processing performance factors. The \ac{hwo} analysis should leverage the \ac{ngrst} \ac{cgi} operations and post-processing concept. This systematic approach should use simulated data cubes to maximize exoplanet detection yield and characterization capabilities.

\subsection{Polarization Differential Imaging}
\focusgrouplead{Max Millar-Blanchaer}

\Ac{pdi} is an observational technique that suppresses starlight by leveraging the relatively weak polarization of stars compared to circumstellar sources of interest (circumstellar disks and planets in reflected light). Polarimetry is a promising approach for directly probing the surface properties of habitable worlds. Earth's ocean glint provides significant polarized signatures that could be used in habitable world characterization. Reflected-light exoplanets may achieve polarization levels exceeding 30\%, as observed in Neptune, Uranus, and Titan.

The \ac{pdi} technique has been successfully applied in both space-based (HST NICMOS) and ground-based (e.g., GPI, \citealt{Macintosh+Graham+Ingraham+etal_2014}; SPHERE,  \citealt{Beuzit+Vigan+Mouillet+etal_2019}; SCExAO,  \citealt{Jovanovic+Martinache+Guyon+etal_2015}\footnote{Instrument acronyms are as follows: \acf{nicmos}, \acf{gpi}, \acf{sphere}, \acf{scexao}}) observations, and typically achieves speckle suppression factors of several hundred. Implementation of PDI requires specialized hardware architecture and typically includes two primary components: a rotatable half-wave plate modulator and a polarizing beamsplitter analyzer that splits incident beams into orthogonally polarized channels imaged either on the same detector (using Wollaston prisms) or separate detectors (using beamsplitter cubes). The technique measures linear Stokes parameters Q and U through systematic modulation sequences, combining multiple exposures into final polarized intensity images that reveal structures with degrees of polarization in the tens of percent regime, as demonstrated in protoplanetary and debris disk observations. \Ac{pdi} observations can also be enabled with only the beamsplitter, in which case the modulator's function must be reproduced in another way — such as the inclusion of a second orthogonally-oriented beamsplitter, as in the case of \ac{ngrst} \ac{cgi}.

For the \ac{hwo}, \ac{pdi} presents both opportunities and technical challenges that require comprehensive modeling to assess performance limits at the 10$^{-10}$ contrast level necessary for Earth-analog detection and characterization. Critical considerations include polarization aberrations introduced by telescope and instrument optics that create polarization-dependent speckles, potentially limiting PDI efficacy compared to ground-based performance where such aberrations have not dominated error budgets. Future simulation frameworks should integrate polarization aberration analysis with realistic speckle field modeling, while also incorporating various instrument architectures to evaluate the impact of the polarizing beamsplitter's location and the method of modulation on overall system performance. 
The modeling approach should leverage forward modeling concepts using knowledge of polarization aberrations to distinguish between polarized speckles and polarized sources (planets and disks), while exploring combined techniques such as RDI+ADI+PDI integration for enhanced systematic error removal. These simulations must include a range of exoplanet polarization properties. 

\subsection{Orbital Differential Imaging}
\focusgrouplead{Hervé Le Coroller}

\Ac{odi} is an advanced high-contrast imaging technique introduced by \citet{Males+Skemer+Close_2013} that exploits the expected Keplerian orbital motion of exoplanets to enhance detection and characterization. The technique operates by utilizing orbital parameters to ``deorbit'' planetary signals across temporal baselines, effectively increasing signal-to-noise ratios. ODI implementations include differential techniques that use orbital motion to disentangle speckle noise from planetary signals, analogous to how field rotation is employed in \ac{adi}, and de-orbiting approaches that stack observations after accounting for predicted orbital motion. 

The K-Stacker algorithm solves the Keplerian equations \citep{LeCoroller+Nowak+Arnold+etal_2015,Nowak+LeCoroller+Arnold+etal_2018,LeCoroller+Nowak+Delorme+etal_2020} to co-add high-contrast imaging observations acquired at different orbital phases, even if the planet has moved significantly on its orbit. This method enables for the first time, the detection of planets that remain undetected in individual epochs (S/N $<$ 3) and whose orbital parameters are unknown. K-Stacker simultaneously retrieves orbital parameters and detects the planet. K-Stacker has re-detected the C1 candidate around Alpha Cen A, a potential super-Earth planet in the habitable zone of this star \citep{LeCoroller+Nowak+Wagner+etal_2022} and provided orbital parameters. This algorithm has also demonstrated the capability to achieve 10$^{-8}$ contrasts at 1~arcsecond separations of $\epsilon$~Eridani with approximately 40 hours of observing time \citep{Tschudi+Schmid+Nowak+etal_2024}. Recent statistical combination techniques, including \ac{mcmc} approaches, integrate multiple independent datasets from high contrast imaging, radial velocity, and proper motion measurements \citep{LeCoroller+Nowak+Arnold+etal_2015,Ruffio+Mawet+Czekala+etal_2018, Thompson+Lawrence+Blakely+etal_2023}. Note that  ODI and Kepler algorithms (i.e., K-Stacker, OCTOFITTER) are complementary techniques that must be applied in two steps in order to reach the higher contrast. ODI is a point spread function (PSF) subtraction method (usable only for fast moving planets), equivalent to ADI, RDI, SDI, or PSF-WFS, to remove instrumental speckles and improve contrast at each individual epoch. However, further contrast gains are achieved in a second step by applying Kepler algorithms (K-Stacker, Octoffiter) to multi-epoch observations (even if the planet remains undetected after PSF subtraction at each individual epoch). For a planet on a Keplerian orbit moving several $\lambda$/D between epochs, ``de-orbiting" with K-Stacker algorithms mitigates residual speckles, thereby enhancing the signal-to-noise ratio (S/N) proportionally to the square root of the total exposure time. 

For the \ac{hwo}, \ac{odi} offers significant operational advantages that could substantially reduce or eliminate requirements for telescope rolls and target slews, effectively making traditional \ac{adi} and \ac{rdi} approaches unnecessary for nearby stellar systems within 5-10 parsecs. The feasibility of \ac{odi} instead of ADI implementation depends critically on planets moving sufficiently rapidly ($>>$1$\times$FWHM-PSF) relative to telescope wavefront evolution timescales during the extended exposure times ($>$~10 hours) required for Earth-analog detection and characterization. An advantage of ODI over ADI is that the smooth exozodi will likely be less problematic. 

Future simulation frameworks for algorithms that use multiple observational visits (e.g., OCTOFITTER, K-Stacker) should model the amplitude and correlation of speckle noise as functions of temporal intervals between epochs, incorporating accurate instrumental modeling that accounts for thermal evolution, pointing errors, and wavelength calibration stability requirements ($<$ 1/10$^{th}$ of cross-correlation peak FWHM). Key parameters for systematic evaluation include spectral resolution effects with realistic wavelength stability errors, range of expected position angle and line-of-sight pointing errors, and long-term instrument stability characteristics. The simulation approach should compare observing scenarios with single-visit strategies (analogous to \ac{ngrst} \ac{cgi} OS11) against multi-epoch approaches utilizing 3-8 visits distributed across expected orbital periods, as may be carried out as part of the \ac{hwo} exoEarth survey. This simulation would enable the quantitative assessment of detection statistics, survey completion timescales, and scientific return optimization for the \ac{hwo} mission architecture.

\subsection{Coherent Differential Imaging}
\focusgroupleads{Susan Redmond, Scott Will}

\Ac{cdi} is a post-processing technique that exploits the mutual incoherence between starlight and planet light to enable speckle suppression without requiring telescope rolls or target slews. The method leverages the fundamental principle that electric fields from planets and stars do not interfere coherently, adding in intensity rather than amplitude, which enables electric field estimation techniques to isolate starlight components for subtraction from science images. CDI implementations encompass both deformable mirror (DM)-based approaches, including pairwise probing (PWP, \citealt{Potier+Mazoyer+Wahhaj+etal_2022}), Kalman filtering, dark-zone maintenance (DZM, \citealt{Redmond+Pueyo+Pogorelyuk+etal_2021}), and speckle area nulling (CDI-SAN, \citealt{Nishikawa_2022}), as well as interferometric techniques such as the self-coherent camera (SCC, \citealt{Baudoz+Boccaletti+Baudrand+etal_2006}) that utilizes spatially-separated pinholes in the Lyot stop to generate interference patterns. Laboratory demonstrations have achieved contrast improvements ranging from factors of 1.5 to 25, with recent experiments reaching raw contrasts of approximately 8$\times$10$^{-8}$ using extended Kalman filter approaches and coherent reference differential imaging (CoRDI, \citealt{Redmond+Pueyo+Por+etal_2024}) techniques that employ PCA-based algorithms for coherent intensity estimation and subtraction.

For the \ac{hwo}, CDI techniques impose minimal additional hardware requirements beyond baseline coronagraph optics, with DM-based methods requiring no supplementary components and SCC approaches needing only modified Lyot stops with additional pinholes coupled with larger optics and oversampled detectors. Critical performance dependencies include available photon flux relative to observatory drift rates and finite stellar diameter effects that can degrade model accuracy for DM-based methods, with usable bandwidths typically limited to 10-20\%. Simulation requirements should address systematic performance modeling as functions of stellar diameter, spectral type, and activity level, as well as imaging filter characteristics, and wavefront control dither amplitude parameters, while incorporating realistic time series of the wavefront error for a given concept of operations. The simulation framework should also include additional incoherent astrophysical sources (such as short-period planets and hot exozodiacal dust) and instrumental artifacts (ghost images from transmissive optics). The CDI processing would benefit from integration with complementary techniques for maximum detection likelihood and false positive mitigation.

\subsection{Spectral Processing}
\focusgrouplead{Jean-Baptiste Ruffio}

Spectral processing encompasses a suite of high-contrast imaging techniques that leverage wavelength-dependent characteristics of exoplanetary atmospheres and stellar point spread functions to enhance detection and characterization capabilities. The methodology includes both low-resolution approaches (R $<$ 300) utilizing broadband spectral features and moderate-to-high resolution techniques (R $\ge$ 1,000) that can resolve the distinct spectral signature of molecular species, including trace molecules. With integral field spectroscopy, classical PSF subtraction techniques would be applicable to any spectral resolution. High-resolution point spectrographs also have the option to rely solely on the high-resolution spectral features through techniques like high-pass filtering and template matching.

For the HWO, spectral processing implementation requires the systematic optimization of spectral resolution parameters to balance signal-to-noise performance, atmospheric inference precision, and observatory design constraints including, for example, detector technology, raw starlight suppression capabilities, telescope diameter, and end-to-end throughput considerations. Critical simulation requirements include comprehensive noise covariance modeling that accounts for speckle residuals and spectral correlation effects and spectral extraction systematics characterization. The trade space analysis should consider science cases for exo-Earths, but also the wider diversity of exoplanets that have different noise limitations. Performance metrics for the trade space analysis should likely include abundance uncertainty limits, exposure time or S/N requirements, and detection yield statistics. Implementation studies should address false positive risk assessment across different spectral resolutions, explore major observatory parameter trades with consistent S/N definitions, and incorporate updated noise budget assumptions reflecting anticipated detector technology capabilities for comprehensive spectral resolution trade-space exploration.

\subsection{PSF subtraction algorithms incorporating telemetry and data to post process, empirical and model-based, including machine learning}
\focusgroupleads{Kellen Lawson, Bin Ren}

Telemetry-aided post-processing represents an advanced approach to stellar \ac{psf} subtraction that leverages observatory telemetry data streams, including wavefront sensing and control (WFSC) measurements, telescope thermal sensors, and ancillary imaging or wavefront control sensors, to enhance speckle suppression performance beyond traditional empirical differential imaging techniques. The methodology encompasses both model-based PSF subtraction approaches that utilize optical models coupled with high-cadence telemetry to synthesize reference PSFs, and machine learning frameworks that incorporate telemetry as auxiliary data streams for pattern recognition and speckle characterization. Recent implementations on JWST data using Space Telescope PSF (STPSF; \citealt{Perrin+etal_2014}) simulations with WFS telemetry have demonstrated the viability of synthetic PSF generation for coronagraphic applications \citep{Greenbaum+etal_2023, Lawson+etal_2024, Feng+etal_2025}, while machine learning methods like Signal-Safe Speckle Subtraction have achieved approximately one magnitude improvement in contrast performance at 1.5 $\lambda$/D separations \citep{Bonse+Gebhard+Dannert+etal_2025}. The use of model-based PSF subtraction offers particular advantages for the \ac{hwo} through reduced observational overheads compared to traditional differential imaging techniques and mitigation of empirical reference limitations including stellar diameter variations, spectral differences, and circumstellar dust contamination. Model-based PSF subtraction also has the potential to remove Poisson noise from the reference PSF.

For the HWO, telemetry-aided post-processing implementation requires comprehensive data acquisition systems that record WFSC measurements, telescope temperature monitoring, and ancillary imaging and wavefront sensing data during all coronagraphic observations to enable both synthetic and empirical PSF subtraction optimization. Critical simulation requirements include assessment of differential imaging technique combinations to maximize Earth-analog detection yield per unit observing time, incorporating realistic overhead estimates and post-processing performance gains within comprehensive observatory yield models. The simulation framework should evaluate computational requirements for real-time post-processing capabilities that enable automated identification of nearby stellar companions or exceptionally bright circumstellar dust to facilitate early sequence termination to improve observational efficiency. Validation studies should utilize extensive testbed datasets incorporating images and telemetry to establish machine learning training foundations, supplemented by transfer learning approaches using on-sky calibration data from \ac{jwst} and \ac{ngrst} observations to demonstrate cross-platform algorithm viability and establish lower-limit performance baselines for advanced post-processing techniques anticipated for \ac{hwo} operations.

\vspace*{-0.5em}\section{Coronagraph Concept of Operations}\vspace*{-0.5em}
\focusgrouplead{Bijan Nemati}

The \ac{hwo} Concept of Operations (ConOps) encompasses the complete operational framework for achieving the primary mission objective of detecting, imaging, and characterizing at least 25 potentially habitable Earth-like exoplanets at unprecedented contrasts exceeding 10$^{-10}$. The ConOps framework integrates mission objectives with operational phases including commissioning, regular science operations, and maintenance protocols, while establishing observation scheduling decisions and data post-processing approaches taking into account autonomous operations and ground-controlled sequences. Related to post processing, high-fidelity data simulation of observing sequences will be used for the systematic evaluation of post-processing technique combinations to maximize exoplanet yield per unit observing time. The \ac{hwo} approach parallels the \ac{ngrst} \ac{cgi} experience, where early formulation phases consider multiple post-processing approaches before converging on specific techniques. 

\begin{figure*}[ht]
\begin{center}
\includegraphics[width=0.9\textwidth]{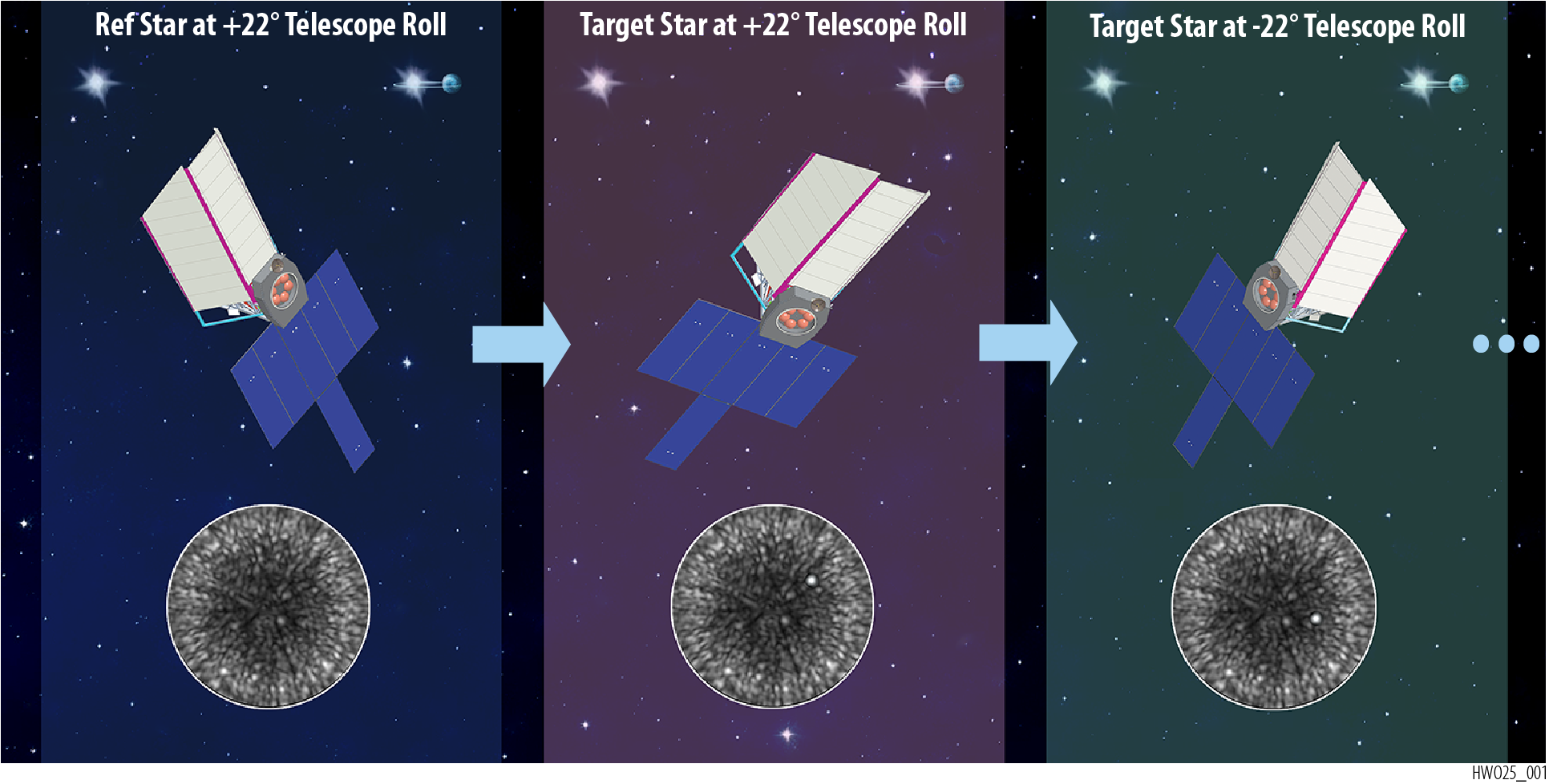}
\caption{\small {\textit{Left:}}The HWO OS1 ConOps uses a reference star for wavefront sensing to expedite the wavefront control sequence. \textit{Center:} After the dark hole is created, the observatory is slewed to a science target with an angular roll of $+$22$^{\circ}$, waits for stabilization, and then an imaging sequence is executed. \textit{Right}: The observatory is rolled to $-$22$^{\circ}$, a 40$^{\circ}$ roll in total, waits for stabilization, and then an imaging sequence is executed. Subsequent rolls and periodic wavefront control maintenance visits to the reference star complete the observing scenario. Simulated data based on HWO integrated modeling results provided by John Krist. 
\label{fig:HWO-OS1}
}
\end{center}
\end{figure*}

\Ac{hwo} Observing Scenario 1 (OS1) was selected as the initial observing sequence as it builds from \ac{ngrst} \ac{cgi} OS11 with minimal modifications to establish baseline performance metrics for differential imaging evaluation (Figure~\ref{fig:HWO-OS1}). The sequence initiates with dark hole generation on a bright reference star, followed by a slew to the target star maintaining identical roll state to minimize thermal transients and wavefront disturbances. The operational timeline implements alternating roll angle observations ($\pm$22$^{\circ}$) for angular differential imaging with periodic reference star revisits for dark hole maintenance, incorporating ground-in-the-loop (GITL) operations where wavefront control probe images are downloaded for ground-based wavefront error analysis and deformable mirror solution computation. OS1 timing parameters include accounting for observatory settling periods following slews and rolls. The scenario serves as the baseline framework for evaluating post-processing performance factors through integrated modeling runs incorporating Structural-Thermal-Optical-Performance (STOP) models that simulate external radiative heat loading variations and telescope optical configuration evolution during the observing sequence.

\section{Post Processing Technology Development}

Post processing gains have been demonstrated using a variety of different approaches on ground and space telescopes, as well as on contemporary high-contrast testbed facilities. 
The Roman CGI currently estimates its coherent on-orbit raw contrast to be 1.4$\times$10$^{-8}$. The estimated gain due to post-processing in the 3-5 $\lambda$/D region ranges from 1.3 to 18.8 depending on noise properties and model uncertainties \citep{Ygouf+Krist+Konomos+etal_2025}. This still leaves a significant gap in applying post-processing methods in the 10$^{-10}$ contrast regime required for HWO.

Post-processing strategies and algorithms are being developed as part of the \ac{hwo} architecture and systems trade studies conducted as part of each Exploratory Analytic Case (EAC). This will ensure that the post-processing approach is consistent with the expected ConOps for each EAC, as well as assumptions about the active wavefront sensing and control system and residual contrast stability. While the \ac{hwo}~OS1 follows an \ac{ngrst} \ac{cgi} approach, additional observing scenarios will be defined to evaluate the range of post processing options discussed in \S~2.

Initial post-processing demonstrations should be performed on simulated or existing data while early testbeds are developed. Once data from the TRL 5 testbeds for starlight suppression, contrast stabilization, and ultra-stable sensing and control are available, they should be incorporated in the post-processing demonstrations consistent with EAC assumptions.

\section{Path Forward}
The \ac{copp} white papers define the state of the art and provide key information to the project regarding coronagraph instrument capabilities, simulations needed, and analysis to be conducted. The post processing approaches should be evaluated using high-fidelity integrated models. While HWO OS1 builds from the mature NGRST OS11, the HWO TMPO would benefit from the definition of a master template of observing scenarios that covers the full range of post processing approaches. The observing scenarios could be used to accurately estimate the post processing impact on yield taking into account important factors such as overheads, duty cycle, and the noise floor. Future testbed facilities should carry out relevant experimental demonstrations to raise the technical maturity.\\

{\bf Acknowledgements.} 

This work was performed within the Concept of Operations and Post-Processing Working Group during the \ac{start} and NASA \ac{tag} phase of the \ac{hwo} \ac{gomap}. The COPP focus groups were supported by a broad range of experts within the community as reported in the white papers. The research was carried out in part through the Jet Propulsion Laboratory, California Institute of Technology, under a contract with the National Aeronautics and Space Administration (80NM0018D0004). RJP was supported by NASA through the CRESST II cooperative agreement CA 80GSFC24M0006. HLC acknowledges this work was supported by the Programme National de Planétologie (PNP) of CNRS-INSU co-funded by CNES and made use of computing facilities operated by CeSAM data center at LAM, Marseille, France. 

\bibliography{author.bib}

{\noindent \bf ACRONYMS}
\begin{small}
\begin{acronym}[parsep=0pt]
\acro{adi}[ADI]{angular differential imaging}
\acro{ao}[AO]{adaptive optics}
\acro{aolp}[AOLP]{angle of linear polarization}
\acro{cdi}[CDI]{coherent differential imaging}
\acro{cgi}[CGI]{Coronagraph Instrument}
\acro{charis}[CHARIS]{Coronagraphic High Angular Resolution Imaging Spectrograph}
\acro{copp}[COPP]{concept of operations and post-processing}
\acro{css}[CSS]{circumstellar scene}
\acro{de}[DE]{differential evolution}
\acro{disnmf}[DI-sNMF]{data imputation using sequential nonnegative matrix factorization}
\acro{dpp}[DPP]{Data Processing Pipeline}
\acro{drp}[DRP]{Data Reduction Pipeline}
\acro{fov}[FOV]{field of view}
\acro{fwhm}[FWHM]{full width at half maximum}
\acro{gomap}[GOMAP]{Great Observatory Maturation Program}
\acro{gpi}[GPI]{Gemini Planet Imager}
\acro{gto}[GTO]{Guaranteed Time Observations}
\acro{hiciao}[HiCIAO]{High-Contrast Coronagraphic Imager for Adaptive Optics}
\acro{hst}[HST]{Hubble Space Telescope}
\acro{hwo}[HWO]{\textit{Habitable Worlds Observatory}}
\acro{hwp}[HWP]{half-wave plate}
\acro{i}[I]{total intensity}
\acro{ir}[IR]{infrared}
\acro{ifs}[IFS]{integral field spectrograph}
\acro{irdis}[IRDIS]{InfraRed Dual-band Imager and Spectrograph}
\acro{iwa}[IWA]{inner working angle}
\acro{jwst}[JWST]{\textit{James Webb Space Telescope}}
\acro{klip}[KLIP]{Karhunen-Lo\`{e}ve Image Projection}
\acro{loci}[LOCI]{Locally Optimized Combination of Images}
\acro{magaox}[MagAO-X]{MagAO-X}
\acro{mcmc}[MCMC]{Markov Chain Monte Carlo}
\acro{ml}[ML]{machine learning}
\acro{mloci}[MLOCI]{Matched LOCI}
\acro{ngrst}[NGRST]{\textit{Nancy Grace Roman Space Telescope}}
\acro{nir}[NIR]{near-infrared}
\acro{nircam}[NIRCam]{Near Infrared Camera}
\acro{nicmos}[NICMOS]{Near Infrared Camera and Multi-Object Spectrometer}
\acro{nmf}[NMF]{Non-negative Matrix Factorization}
\acro{opd}[OPD]{optical path difference}
\acro{odi}[ODI]{orbital differential imaging}
\acro{pca}[PCA]{principal component analysis}
\acro{pdi}[PDI]{polarized differential imaging}
\acro{pi}[PI]{polarized intensity}
\acro{piaacmc}[PIAACMC]{phase-induced amplitude apodization complex mask coronagraph}
\acro{ppd}[PPD]{protoplanetary disk}
\acro{psf}[PSF]{point spread function}
\acro{rdi}[RDI]{reference star differential imaging}
\acro{sb}[SB]{surface brightness}
\acro{scexao}[SCExAO]{Subaru  Coronagraphic  Extreme  Adaptive  Optics}
\acro{sdi}[SDI]{spectral differential imaging}
\acro{sed}[SED]{spectral energy distribution}
\acro{snr}[SNR]{signal-to-noise ratio}
\acro{snre}[SNRE]{signal-to-noise per resolution element}
\acro{spf}[SPF]{scattering phase function}
\acro{sphere}[SPHERE]{Spectro-Polarimetric High-contrast Exoplanet REsearch instrument}
\acro{start}[START]{Science, Technology, Architecture Review Team}
\acro{tag}[TAG]{Technical Assessment Group}
\acro{vampires}[VAMPIRES]{Visible Aperture Masking Polarimetric Imager for Resolved Exoplanetary Structures}
\acro{vapp}[vAPP]{vector Apodizing Phase Plate}
\acro{vlt}[VLT]{Very Large Telescope}
\end{acronym}
\end{small}

\end{document}